\documentclass[preprint,12pt]{elsarticle}

\usepackage{amssymb}
\usepackage{amsmath}
\usepackage{tabularx}

\usepackage{booktabs}   
\usepackage{array}      
\usepackage{hyperref}

\journal{arXiv} 

\begin{document}

\begin{frontmatter}


\affiliation[scls]{organization={Shanghai Soong Ching Ling School}, 
            addressline={No. 2 Ye Hui Road}, 
            city={Shanghai},
            postcode={201703}, 
            country={China}}

\affiliation[cwi]{organization={China Welfare Institute Children's Palace}, 
            addressline={No. 64 Yan'an West Road, Jing'an District}, 
            city={Shanghai},
            postcode={200003}, 
            country={China}}

\author[scls]{\texorpdfstring{Tong Hu\corref{cor1}}{Tong Hu}}
\ead{tong.hu@scls-sh.org.cn}

\author[cwi]{Songzan Wang}

\cortext[cor1]{Corresponding author}

\title{From Generation to Adaptation: Comparing AI-Assisted Strategies in High School Programming Education}


\begin{abstract}
This exploratory case study investigated two contrasting pedagogical approaches for LCA-assisted programming with five novice high school students preparing for a WeChat Mini Program competition. In Phase 1, students used LCAs to generate code from abstract specifications (From-Scratch approach), achieving only 20\% MVP completion. In Phase 2, students adapted existing Minimal Functional Units (MFUs)---small, functional code examples---using LCAs, achieving 100\% MVP completion. Analysis revealed that the MFU-based approach succeeded by aligning with LCA strengths in pattern modification rather than \textit{de novo} generation, while providing cognitive scaffolds that enabled students to navigate complex development tasks. The study introduces a dual-scaffolding model combining technical support (MFUs) with pedagogical guidance (structured prompting strategies), demonstrating that effective LCA integration depends less on AI capabilities than on instructional design. These findings offer practical guidance for educators seeking to transform AI tools from sources of frustration into productive learning partners in programming education.

\end{abstract}


\end{frontmatter}



\section{Introduction}
\label{sec1}

Artificial Intelligence (AI) is increasing recognized as a key driver of educational innovation\cite{tan_artificial_2025}. Large Language Model Coding Agents (LCAs) can significantly enhance learning in Computer Science education by offering students personalized guidance on key activities such as coding, debugging, and comprehending fundamental principles\cite{chu_llm_2025}. These models promise to democratize education, personalize learner support\cite{zhai_review_2021}, and accelerate complex software development, lowering the barrier for students. Recent studies and reviews, however, reveal that while AI applications in education span from adaptive learning\cite{li_bringing_2024} to intelligent assessment\cite{poretschkin_guideline_2023}, robust empirical evidence for effective classroom integration remains limited. Recent studies suggest that while LCAs can boost immediate task performance, this does not always translate into statistically significant learning gains for novice users\cite{kazemitabaar_studying_2023}.

Within educational contexts, much of the current enthusiasm for LCAs stems from the belief that they enable even novice coders to build sophisticated applications from scratch – a “0-to-1” leap.\cite{Lovable_Docs_IdeaToApp} Early classroom trials, including the one that motivates this study, paint a more nuanced picture. When learners rely on LCAs to generate entire projects from abstract specifications, they often struggle to craft precise prompts\cite{becker_programming_2022}---a skill now recognized as crucial for effective AI interaction\cite{white_prompt_2023}, debug subtle code defects, and weave AI-generated fragments into a coherent whole. Recent research with novice programmers found that while AI tools can boost programming efficiency and lighten cognitive workload, they do not necessarily enhance learning satisfaction or self-efficacy without proper pedagogical support\cite{gardella_performance_2024}.

This explanatory case study addresses this gap by investigating the effectiveness of a specific scaffolded pedagogical approach for LCA-assisted project development by novice high school coders. This study contrasts a From-Scratch (or "0-to-1") strategy, where students used LCAs to build software based on their initial plans, with an alternative strategy that leverages to adapt and extend existing "Minimal Functional Units" (MFUs)---defined for this study as small, functional applications that demonstrate core features relevant to target projects. We term this the MFU-based (or "1-to-100") approach. The primary purpose of this research is to evaluate the differential outcomes of these two approaches, understand the underlying factors related to LCA characteristics and human-AI collaboration that contribute to these differences, and derive initial pedagogical insights.

The central argument of this paper is that for novice high-school coders who possess basic programming skills but lack prior hands-on development experience, the MFU-based approach proved markedly more effective than the From-Scratch strategy in producing functional projects. This study is guided by the following research questions:

\begin{enumerate}
    \item[\textbf{RQ1:}] How do project outcomes (e.g., Minimum Viable Product (MVP) completion rates) and student development experiences differ when novice high school coders use LCAs under a \textit{From-Scratch} versus an \textit{MFU-based} pedagogical strategy?
    \item[\textbf{RQ2:}] What factors related to LCA tool performance and human-LCA interaction dynamics contribute to these observed differences?
    \item[\textbf{RQ3:}] What are the key pedagogical implications for effectively integrating LCAs into K-12 programming education based on these findings?
\end{enumerate}

Given its exploratory nature and the small-sample, single-site case context (detailed in Chapter 2), this study does not aim to yield broadly generalizable conclusions. Rather, its primary contribution lies in offering heuristic value and practical insights by presenting a detailed account of a pedagogical intervention and its outcomes. It seeks to provide actionable guidance for K-12 educators navigating the integration of LCAs and to encourage further research into human-AI pedagogy. 

The paper is structured as follows: Section 2 details the research methodology; Section 3 presents the core findings; Section 4 discusses their significance, theoretical connections, pedagogical implications, limitations, and future research; Section 5 offers a concise conclusion.

\section{Methods}
\label{sec2}

This exploratory case study investigated the effectiveness of different LCA-assisted development strategies for novice high school coders. The study compared two approaches: From-Scratch generation versus adaption of existing Minimal Functional Units (MFUs).

\subsection{Participants and Context}
\label{subsec1}

Five students (Grades 10-12) from a private high school in Shanghai participated in this study while preparing for the 3rd WeChat Mini Program Innovation Challenge. All participants had completed AP Computer Science A but lacked prior WeChat Mini Program development experience. The study took place during two evening coding sessions in April 2025, each lasting approximately 3 hours.

\subsection{Materials and Tools}

Students used Claude 3.7 Sonnet through Windsurf on their own laptop. MFUs were sourced from the official WeChat Mini Program example library. The study received ethical approval from the school, with informed consent from participants and guardians.

\subsection{Procedure}
Session 1: Students attempted to create WeChat Mini Programs from scratch using LCA-generated code based on their project specifications.

Session 2: Following limited success in Session 1, students adopted an MFU-based approach, selecting relevant examples and using LCAs to adapt them for their projects.

\subsection{Data Collection and Analysis}
Data consisted primarily of instructor field notes recorded during and immediately after each session, documenting strategies, student behaviors, technical challenges, and MVP completion. An MVP was defined as a functioning Mini Program demonstrating at least one core feature.

Analysis involved systematic comparison of the two sessions, identifying patterns in student-LCA interactions and factors contributing to differential outcomes. The instructor’s dual role as researcher-teacher was acknowledged as both providing insider insights and potential bias.

\section{Findings}

This section presents comparative findings from two coding sessions with contrasting pedagogical approaches. 

\subsection{Phase 1: From-Scratch Approach}

During the first session, five teams attempted to develop WeChat Mini Programs using LCA-generated code based on their project specifications. Table 1 summarizes the outcomes.

\begin{table}[htbp]
\centering
\caption{Phase 1 Project Outcomes and Technical Challenges ("From-Scratch" Approach)}
\label{tab:phase1_outcomes}

\begin{tabularx}{\textwidth}{>{\raggedright\arraybackslash}X >{\raggedright\arraybackslash}X >{\raggedright\arraybackslash}X l}
\toprule
\textbf{Project Name} & \textbf{Core Feature} & \textbf{Key Technical Bottleneck} & \textbf{Outcome} \\
\midrule
Myopia Simulator      & Real-time vision simulation & Inability to activate or utilize the phone's camera function & Stalled          \\
Campus Tree Hole      & Anonymous campus blog platform & Persistent errors during Mini Program cloud service access & Stalled          \\
AI Campus Assistant   & AI-powered school tools & Functional in simulator but failed on actual device via QR code scan & Stalled          \\
Carbon Footprint      & Transportation option tracking & None reported; core functionality achieved & MVP Achieved     \\
WhereToGo             & AI route recommendations & Could not load the application's homescreen & Stalled          \\ 
\bottomrule
\end{tabularx}
\end{table}

Only one team (20\%) achieved a functional MVP. Instructor observations documented frequent student requests for debugging help, with LCA outputs often requiring substantial modification. Common issues included syntactically correct but non-functional code and missing architectural context specific to WeChat Mini Programs.

\subsection{Pedagogical Shift}

The limited success in Phase 1 prompted a strategic revision. Observing that the LCA performed better when modifying existing code rather than generating from scratch, the instructor introduced MFUs---small, functional applications from the official WeChat Mini Program library that demonstrates features relevant to student projects---as starting points for Phase 2.

\subsection{Phase 2: MFU-based Approach}

Teams selected MFUs containing relevant technical components, though the MFUs’ original purposes differed entirely from student projects. The Myopia Simulator team, previously blocked by camera issues, chose an object classification mini-program that demonstrates video streaming, extracting only the camera functionality while discarding the classification features. AI Campus Assistant adapted a basic chatbot MFU, retaining its AI interaction framework but completely rebuilding its purpose from general conversation to educational tools like Multiple-Choice Question generation and lesson planning.
Students’ prompts evolved from single requests to multi-step structured prompts. Table 2 illustrates this shift using the Myopia Simulator project as an example.

\begin{table}[htbp]
\centering
\caption{Student Prompts: Phase 1 vs. Phase 2}
\label{tab:prompt_comparison}
\begin{tabularx}{\textwidth}{l p{3.5cm} X} 
\hline
\textbf{Phase} & \textbf{Prompt Strategy} & \textbf{Example} \\ \hline
Phase 1 & Single, abstract request & "Create a mini program that simulates myopia with camera functions" \\ \hline
Phase 2 & Multi-step analytical process & Step 1: "Analyze this object classifier MFU and list all features" \newline 
Step 2: "Compare with my vision simulator requirements. What to keep/remove?" \newline
Step 3: "Remove classification, keep camera module, add blur filters for myopia simulation" \\
\hline
\end{tabularx}
\end{table}

All five teams (100\%) achieved functional MVPs in Phase 2. Instructor notes documented several key improvements across technical, affective, and strategic dimensions:

\begin{itemize}
    \item \textbf{Technical:} Students required fewer debugging requests and achieved faster feature implementation.
    \item \textbf{Affective:} Students displayed greater confidence and maintained optimistic attitudes throughout the session.
    \item \textbf{Strategic:} Students formulated more structured prompts that referenced specific MFU components, demonstrating a more advanced interaction with the LCA.
\end{itemize}

\subsection{Summary}

Phase 1’s From-Scratch approach yielded 20\% MVP completion amid substantial technical challenges. Phase 2’s MFU-based strategy achieved 100\% completion with improved efficiency and student confidence. These outcomes suggest that successful LCA integration depends on alignment between pedagogical approach, tool capabilities, and student needs---themes explored further in the discussion.

\section{Discussion}

The stark contrast between 20\% and 100\% MVP completion rates illuminates a fundamental insight: LCA success in education depends less on AI capabilities than on pedagogical design. This study’s dual-scaffolding model---technical MFUs plus structured pedagogy---offers a framework for transforming AI from obstacle to enabler. 

\subsection{Why MFUs Work}

MFUs succeed by aligning with both human cognition and AI architecture. For students, they provide concrete anchors in the “adjacent possible”---close enough to understand, far enough to require creative adaptation. For LCAs, they offer what these models need most: context-rich patterns to modify rather than void-to-code generation tasks. This alignment is supported by recent research showing that LLMs perform significantly better when provided with relevant code context\cite{Google_DeeperInsightsRAG_2025}. 

This alignment explains our results. When students prompt “create a camera function”, LCAs must infer entire architectures. When students prompt “adapt this camera module for myopia simulation”, LCAs excel at pattern transformation---their core strength. The pedagogical shift from “imagine then generate” to “analyze then adapt” fundamentally changes the human-AI dynamic. 

\subsection{Rethinking Teacher Roles}

The study also adds to evidence that teacher's role in AI-mediated classrooms is shifting from content transmitter to collaboration orchestrator. Recent Teacher-AI Collaboration (TAC) models position educators as orchestrators who leverage AI to foster higher-order thinking and rich peer interaction\cite{holstein_co-designing_2019, kim_types_2024}. Crucially, process scaffolding outweighs content delivery. A study by Matsuda \cite{matsuda_effect_2020} showed that meta-cognitive scaffolds on how to learn yield larger gains than cognitive scaffolds on what to learn, and finding further supported by the advantages of AI-mediated process support\cite{Wang04032025}. By framing the LCA as a thought partner rather than an answer machine, teachers cultivate students’ AI literacy and critical thinking, skills essential for an AI-augmented future.

\subsection{Limitations and Future Research}

This exploratory study has several limitations. The small sample (n=5) of motivated high school students from a single private school in Shanghai limits generalizability. The instructor’s dual role as researcher and teacher may introduce bias, while the brief study duration prevented long-term follow-up. Additionally, the study relied predominantly on instructor observations and completion rates, lacking systematic measures such as cognitive load assessments, self-efficacy scales, or structured student feedback. While the sequential design prevents pure causal claims, the pronounced performance gap nonetheless offers preliminary evidence that MFU-based scaffolding warrants further controlled investigation.

Future research should address these gaps through controlled experiments comparing MFU-based and traditional approaches across diverse educational contexts---varying grade levels, programming backgrounds, and school types. Additionally, future research should combine quantitative measures of cognitive load and self-efficacy with qualitative insights into student experiences. The critical question of skill transfer---whether MFU-based learning enables independent programming---requires longitudinal investigation. Additionally, research should explore optimal MFU selection criteria and the evolution of scaffolding needs as LCA technology advances.

Despite these limitations, this study provides actionable guidance for educators integrating LCAs into programming curricula and establishes a testable framework for human-AI pedagogical models in K-12 contexts. 

\section{Conclusion}

The stark contrast between 20\% and 100\% MVP completion rates illuminates a fundamental insight: LCA success in education depends less on AI capabilities than on pedagogical design. MFU-based scaffolding enabled all five high-school teams to complete WeChat Mini Program MVPs, versus only 20\% using \textit{de novo} generation. This dual-scaffolding approach---combining technical MFUs with pedagogical guidance---demonstrates that successful LCA integration requires deliberate instructional design, not merely AI access. The framework developed here offers practical guidance for educators seeking to transform AI tools from barriers into bridges in programming education.

\bibliographystyle{apalike} 
\bibliography{LCA}



\end{document}